# *p*-type co-doping effect in (Ga,Mn)As: Mn lattice location vs. magnetic properties

Chi Xu[1,3,†], Chenhui Zhang[2,†], Mao Wang[1,3], Yufang Xie[1,3], René Hübner[1], René Heller[1], Ye Yuan[1,2,*], Manfred Helm[1,3], Xixiang Zhang[2], and Shengqiang Zhou[1]

[1]Helmholtz-Zentrum Dresden-Rossendorf, Institute of Ion Beam Physics and Materials Research, Bautzner Landstrasse 400, 01328, Dresden, Germany

[2]King Abdullah University of Science and Technology (KAUST), Physical Science and Engineering Division (PSE), Thuwal 23955-6900, Saudi Arabia

[3]Technische Universität Dresden, 01062, Dresden, Germany

In the present work, we perform a systematic investigation on *p*-type co-doping in (Ga,Mn)As. Through gradually increasing Zn doping concentration, the hole concentration increases, which should theoretically lead to an increase of the Curie temperature ($T_C$) according to the *p-d* Zener model. Unexpectedly, although the film keeps its epitaxial structure, both $T_C$ and the magnetization decrease. The samples present a phase transition from ferromagnetism to paramagnetism upon increasing hole concentration. In the intermediate regime, we observe a signature of antiferromagnetism. By using channeling Rutherford-Backscattering Spectrometry and Particle-Induced X-ray Emission, the substitutional Mn atoms are observed to shift to interstitial sites, while more Zn atoms occupy Ga sites, which explains the observed behavior. This is also consistent with first-principles calculations, showing that the complex of substitutional Zn and interstitial Mn has the lowest formation energy.

* Email: ye.yuan@kaust.edu.sa

# I. Introduction

The III-V dilute ferromagnetic semiconductors (DFS), in which a fraction of the cations is substitutionally replaced by 3$d$ transition metal atoms, behave as both semiconductors and ferromagnets [1-6]. In DFS, the ferromagnetism is mediated by itinerant holes, which has been successfully explained by the Zener mean-field model [3, 7, 8]. For the most canonical DFS (Ga,Mn)As, the Curie temperature ($T_C$) is determined by the effective Mn concentration (Mn$_{eff}$) and the hole concentration $p$ as $T_C \propto$ Mn$_{eff}p^{1/3}$ [3, 7, 9]. Another example solidifying the importance of holes is given by $p$-type nitrogen-doped (Zn,Mn)Te epitaxial layers, in which the ferromagnetic coupling with a $T_C$ of 2.4 K emerges, once the hole concentration approaches 1.5×10$^{19}$ cm$^{-3}$ together with an effective Mn concentration of 2.7% [10, 11]. Thus, intentionally increasing the hole concentration is always placed as a central task to increase $T_C$ and manipulate the magnetic properties in DFS. This has been attempted by electrical gating [4, 8, 12-15] as well as acceptor co-doping [10, 11]. However, the threshold Mn concentration for the presence of global ferromagnetism is several orders of magnitude higher than its equilibrium solid solubility in III-V semiconductors, therefore prohibiting most preparation ways for conventional co-doping of epitaxial thin films [1, 16]. Nevertheless, Devillers et al. successfully obtained high-quality Mg-doped (Ga,Mn)N epitaxial films by metalorganic vapor phase epitaxy (MOVPE) [17]. Unfortunately, during the growth process, Mg $p$-dopant atoms diffuse and approach the Mn atoms to finally forming Mn-Mg dimers, which largely quenches the Mn moment [17]. Thus, a novel preparation of co-doping acceptor elements in III-Mn-V DFS is of great importance.

In the present work, we have systematically increased the carrier concentration in the canonical DFS (Ga,Mn)As through Zn co-doping by using ion implantation combined with pulsed laser melting and performed investigations on structural, electronic and magnetic properties of the prepared samples. We unexpectedly find that, while keeping the epitaxial structure, both $T_C$ and magnetization decrease upon increasing the hole concentration by Zn co-doping. Rutherford Backscattering Spectroscopy (RBS) and Particle-Induced X-ray Emission (PIXE) results and *ab initio* calculations consistently show, that the substitutional doping of Zn energetically drives Mn atoms into interstitial sites, which results in the reduction of the magnetization in the layer. Our results indicate, that the way of co-doping for introducing holes into the DFS system to increase $T_C$ is quite difficult and a solution to avoid Mn interstitialization during the preparation is urgently required.

## II. Experimental methods

### A. Sample Preparation

The (Ga,Mn)As and Zn co-doped films for this study were prepared by Mn and Zn ion co-implantation into semi-insulating GaAs wafers followed by pulsed laser melting (PLM). The implantation fluence was $8\times10^{15}$ cm$^{-2}$ and the implantation energy was 100 keV for Mn. The corresponding implantation energy for Zn was carefully selected to 120 keV to ensure a complete overlap of the Mn and Zn distributions. Both depth profiles of Mn and Zn atoms in doped GaAs were calculated by the Stopping and Range of Ions in Matter (SRIM) code [18] and the thickness is two times the longitudinal straggling ($\Delta R_P$), i.e. $\Delta R_P$ for Mn and Zn are both around 30 nm, thus the thickness of the doped layers is 60 nm accordingly. Similarly, the peak Mn concentration $x$ in Ga$_{1-x}$Mn$_x$As is estimated to be 4.7% according to SRIM. Four different Zn implantation fluences were chosen to realize a gradual increase of the co-doping density in (Ga,Mn)As. Together with the sample that is only doped with Mn, the samples are named as Zn-0, Zn-1, Zn-2, Zn-4, and Zn-8 according to the Zn implantation fluences of 0, $1\times10^{15}$, $2\times10^{15}$, $4\times10^{15}$, and $8\times10^{15}$ cm$^{-2}$, respectively. A commercial XeCl pulsed excimer laser (Coherent ComPexPRO201, wavelength $\lambda$ = 308 nm and pulse duration $t$ = 30 ns) was used for the PLM treatment which was carried out after the successive implantations of Mn and Zn ions. The energy density of 0.3 J/cm$^2$ was adopted to get the best epitaxial structure as previously reported [19]. It is worth noting that such an ultrafast pulsed laser annealing causes a non-equilibrium recrystallization, which is different from the equilibrium annealing before [20]. To remove the Mn-rich oxide layer formed during the PLM process, all samples were immersed in a (1:10) HCl solution at room temperature for 1 hour [21]. Other groups have also used such an approach combining ion implantation and PLM to prepare (Ga,Mn)As and (Ga,Mn)P [22-24]. Chen et al. also tried to use ion irradiation to recrystallize Mn implanted GaAs [25].

### B. Sample Characterization

The magnetic properties of the samples were studied by using a Quantum Design MPMS 3 Superconducting Quantum Interference Device - Vibrating Sample Magnetometer (SQUID-VSM). For the magnetization curves at 5 K, the diamagnetic signal contribution of the GaAs substrate was subtracted by the room-temperature $M$-$H$ curves. For the thermo-remanent magnetization (TRM) measurements, the sample was firstly cooled down to 5 K under an external field of 2 kOe, then the field was reset to zero by a superconducting magnet quench, and data was collected during warming. For all magnetic measurements, the magnetic field was applied along the in-plane $[1\bar{1}0]$ direction which is the magnetic easy axis for (Ga,Mn)As. The magneto-transport measurements for all samples were performed with a magnetic field applied perpendicularly to the film plane in a Lake Shore Hall Measurement System using the van-der-Pauw geometry. The electrical contact was made by silver paste. The structural properties of the PLM-treated samples were studied by Rutherford Backscattering Spectrometry/Channeling (RBS/C) which was performed with a collimated 1.7 MeV $He^+$ beam from the Rossendorf van-de-Graff accelerator with 10-20 nA beam current at a backscattering angle of 170°. RBS and PIXE channeling spectra were collected by aligning the sample to make the impinging $He^+$ beam parallel with the [011] and [001] axis of the substrate. During collecting the random spectra, the GaAs sample was tilted by 4° and rotated around its surface normal. The random spectra were High-resolution transmission electron microscopy (TEM) analysis was performed using an image-$C_s$-corrected Titan 80-300 microscope (FEI) operated at an accelerating voltage of 300 kV. For some results the error bars are given, but they only include the statistical error.

### C. First-principles Calculations

Our first-principles calculations were performed using spin-polarized density functional theory (DFT), as implemented in the Vienna *ab initio* simulation package (VASP) code [26]. The projector-augmented wave (PAW) potentials [27] were adopted to describe the core electrons, and the generalized gradient approximation (GGA) of Perdew, Burke, and Ernzernhof (PBE) [28] was used for exchange–correlation functional. In order to simulate the experimental dopant concentration, a zinc-blende GaAs unit cell was extended to a 2×2×3 supercell, then one Mn atom replaced one Ga atom and one additional Zn atom was placed on an interstitial site. So, the whole structure contained 48 As atoms, 47 Ga atoms, one Mn atom and one Zn atom. Both the lattice parameter and atomic positions were relaxed to a force tolerance of 0.01 eV/Å, while the cut-off energy was set to 360 eV. The Brillouin zone sampling was done using a 6×6×4 Monkhorst-Pack grid.

## III. Results and discussion

### A. Structure Characterization

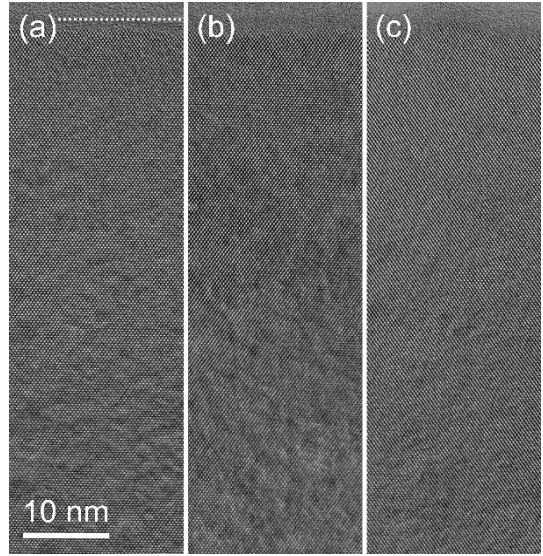

FIG. 1 Cross-sectional high-resolution TEM micrographs of samples (a) Zn-0, (b) Zn-2, and (d) Zn-8. The sample surface is exemplarily marked by a dashed line in (a).

The microstructure of the Mn- and Zn-co-doped samples was investigated by cross-sectional high-resolution TEM. Representative micrographs covering the whole implanted layer thickness of 62 nm (as calculated by SRIM) are displayed in Fig. 1. For all PLM-treated (Ga,Mn)As:Zn samples, epitaxial recrystallization has been achieved. Such recrystallization during PLM in the present case is worth mentioning, since it is hard to obtain highly epitaxial materials alloyed with the group III or V elements [29, 30] or co-doped with acceptors [31] through low-temperature molecular beam epitaxy (LT-MBE).

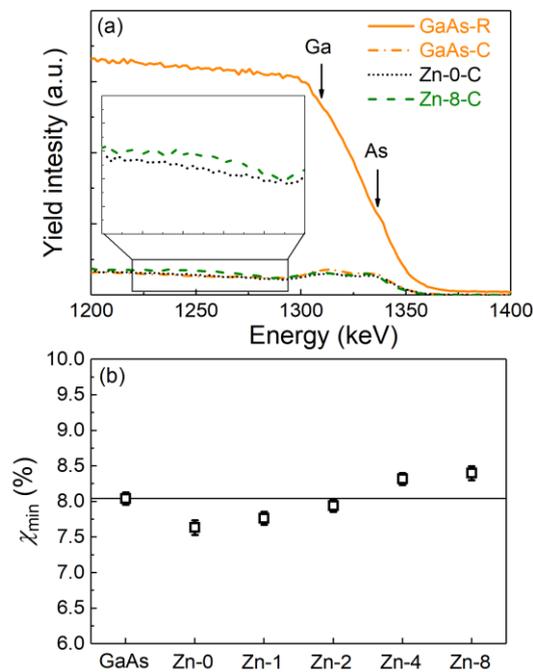

FIG. 2 (Color online) (a) Channeling [110] Rutherford backscattering spectra of samples Zn-0 (dot) and Zn-8 (dash). A pure GaAs wafer (dash dot) is placed as a reference for comparison as well as the random spectra (solid line). The insert shows the sign of interstitial Mn atoms in sample Zn-8. (b) Calculated minimum backscattering yield $\chi_{min}$ [32] for all samples together with the pure GaAs for the reference.

The recrystallization of the PLM-treated (Ga,Mn)As:Zn layers is further investigated by RBS/C. Figure 2(a) shows representative RBS/C spectra of the PLM-treated (Ga,Mn)As:Zn samples with lowest and highest Zn concentrations together with a virgin GaAs wafer as a reference for comparison. The channeling spectra are recorded by aligning the GaAs [011] axis, which is more sensitive to interstitial Mn impurities [31, 33, 34]. For the random spectra, Ga and As signals are both prominent, whereas the Mn signal is not visible due to its low concentration of only several percent and its overlap with the Ga and As signals. From the RBS-channeling signal, the PLM-treated (Ga,Mn)As:Zn layers show the channeling effect with the surface peaks located around 1320 and 1330 keV referring to Ga and As, respectively. In addition, a minimum backscattering yield $\chi_{min}$ of around 8.0% can be estimated (defined as the ratio of the aligned to the random yields [32] in the energy range 1275-1300 keV), which is comparable to that of the virgin GaAs substrate, as shown in Fig. 2(b). In summary, both RBS/C and high-resolution TEM analysis confirm an epitaxial regrowth of the (Ga,Mn)As:Zn layers during the PLM treatment.

### B. Magneto-transport properties

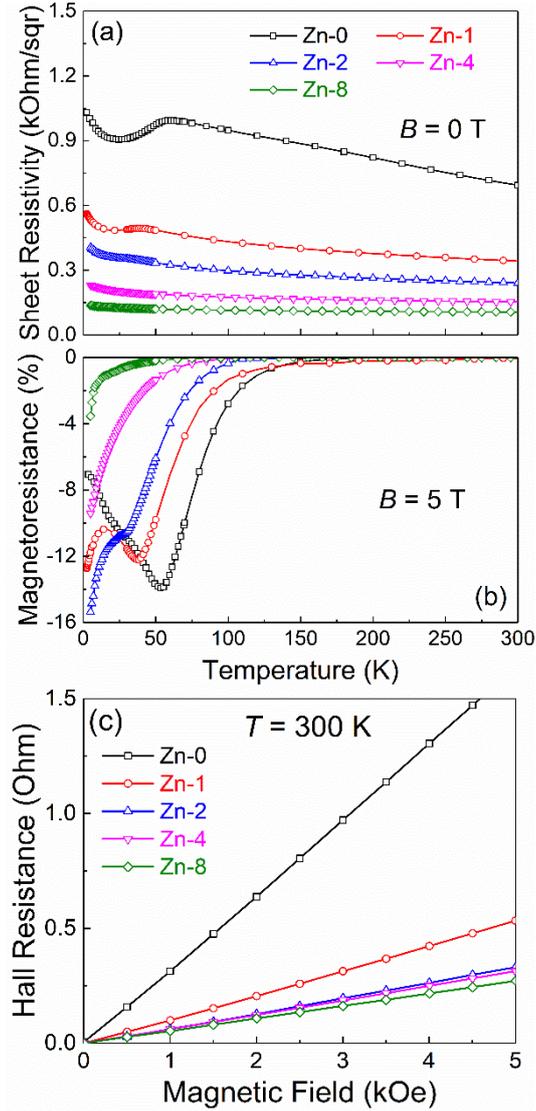

FIG. 3 (Color online) (a) Temperature-dependent sheet resistance at zero field and (b) magnetoresistance under a 5 T field for (Ga,Mn)As and (Ga,Mn)As:Zn samples. (c) The magnetic-field-dependent Hall resistance at room temperature of all samples.

To confirm the hole-doping effect in Zn-co-doped samples, electrical properties were investigated and results are shown in Fig. 3. As displayed in Fig. 3(a), upon increasing temperature from 5 K to 60 K which is around the $T_C$, the resistance of the Zn-0 sample presents metallic behavior, which is consistent with the typical ferromagnetic (Ga,Mn)As sample. Moreover, as expected, with increased Zn doping, the sheet resistance gradually reduces, indicating that the carrier concentration is indeed increased by the Zn dopants, quantitatively confirmed by the Hall resistance, as shown in Fig. 3(c). It is worth mentioning that due to the measurement at room temperature, which is far above $T_C$, the anomalous Hall effect is minimized and the ordinary Hall effect dominates. Accordingly, all magnetic-field-dependent Hall resistance curves show positive slopes, which suggests the *p*-type conducting behavior in all samples.

Additionally, according to the Hall resistance slopes, hole concentrations are calculated to be $1.6\times10^{20}$, $4.9\times10^{20}$, $7.9\times10^{20}$, $8.4\times10^{20}$, and $9.6\times10^{20}$ cm$^{-3}$ for samples Zn-0, Zn-1, Zn-2, Zn-4, and Zn-8, respectively, again proving the successful substitutional Zn doping. The increase of hole concentration introduced by Zn doping was also confirmed in LT-MBE grown (Ga,Mn)As [35-37]. In the as-grown samples [31], the authors also found that although the hole concentration is higher in Zn co-doped sample the Curie temperature is lower compared with un-codoped (Ga,Mn)As. Upon post-growth annealing at low temperatures, the authors postulate the formation of GaAs:Zn and MnAs/Zn-Mn-As complexes in Zn co-doped samples [31].

The magnetoresistance (MR) measurement was performed here as well, because it carries fruitful magnetic and electrical information of the sample, and the results are displayed in Fig. 3(b). Herein, the temperature-dependent magnetoresistance is calculated as $\mathrm{MR} = \frac{\rho(5\mathrm{T})-\rho(0)}{\rho(0)} \times 100\%$, where $\rho(5\mathrm{T})$ and $\rho(0)$ are the resistances from 5 K to 300 K at an external magnetic field of 5 T and 0 T, respectively. For samples Zn-0, Zn-1 and Zn-2, the temperature-dependent MR shows a behavior typical for ferromagnetic semiconductors: (i) A negative MR is observed, resulting from the reduction of scattering between spin-polarized holes and Mn local moments, when an external magnetic field is applied. (ii) A maximum negative MR appears at around $T_\mathrm{C}$, which can be understood as the scattering of carriers by thermally magnetic spin fluctuation via exchange interaction, which has been observed in magnetic metals and semiconductors [36, 38]. However, with increased Zn doping, this peak shifts to lower temperature (together with decreased strength), until it totally vanishes for the samples Zn-4 and Zn-8, indicating that increasing the Zn concentration probably results in the reduction of $T_\mathrm{C}$ [9, 38] and even the absence of global ferromagnetism for samples Zn-4 and Zn-8.

### C. Magnetic properties vs. Zn doping

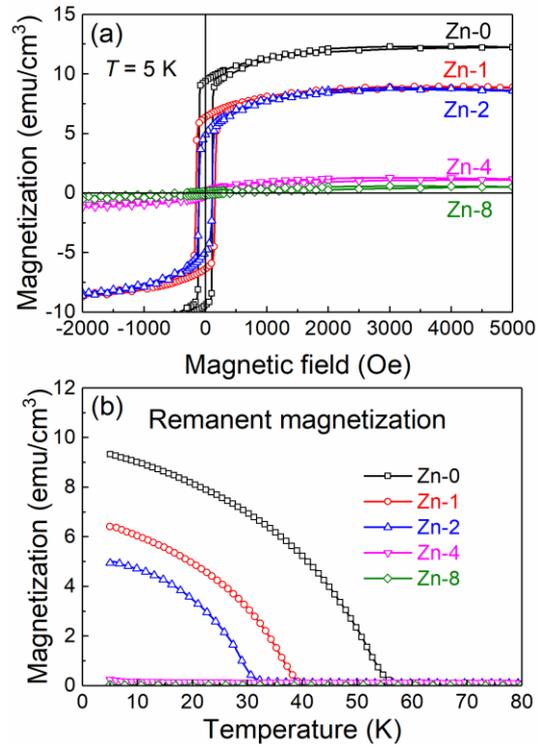

FIG. 4 (Color online) (a) Magnetization as a function of magnetic field at 5 K when magnetic field is applied along the $[1\bar{1}0]$ direction. (b) Temperature-dependent remanent magnetization under zero field after cooling down process under a field of 2 kOe.

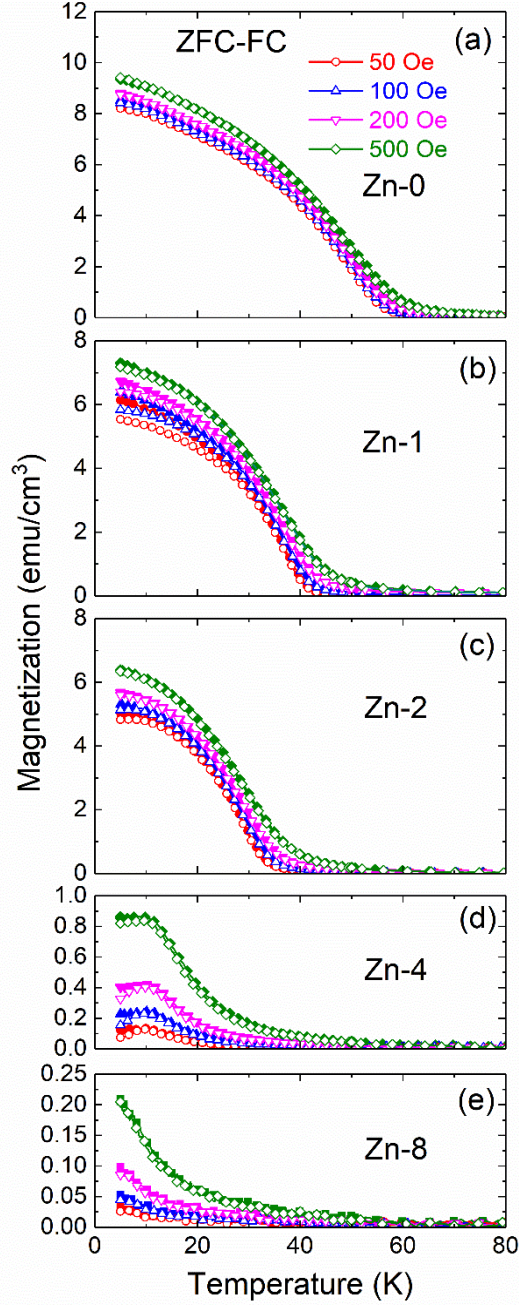

FIG. 5 (Color online) Temperature-dependent magnetization after zero-field-cooling (closed symbol) and field-cooling (open symbol) procedure of samples (a) Zn-0, (b) Zn-1, (c) Zn-2, (d) Zn-4, and (e) Zn-8 under fields of 50 Oe (circles), 100 Oe (up-triangles), 200 Oe (down-triangles), and 500 Oe (diamonds). A peak around 10 K can be observed in (d) as the sign of the possible antiferromagnetic coupling in sample Zn-4.

For detailed investigation of the magnetic properties, particularly the dependence of magnetization and $T_C$ on Zn doping, magnetic field dependent magnetization and temperature-dependent thermo-remanent magnetization (TRM) measurements were carried out for all samples, as shown in Fig. 4. In Fig. 4(a), after subtracting the diamagnetic background of GaAs substrate, the $M$-$H$ curves at 5 K of all samples with the field along the [1$\bar{1}$0] direction are present. For samples Zn-0, Zn-1 and Zn-2, square-like hysteresis loops are seen, which indicates that the ferromagnetism is preserved. When the magnetic field is larger than 2 kOe, the magnetization becomes saturated and defined as the saturation magnetization $M_S$. For samples Zn-4 and Zn-8, both remanent magnetization and saturation magnetization dramatically reduce or even vanishes, indicating the disappearance of ferromagnetism. Before TRM measurement, all samples were cooled down under a field of 2 kOe, which is high enough to saturate the ferromagnetic moment from 300 K to 5 K. Subsequently, the magnetic field was removed by magnet reset operation, making sure that the residual magnetic field is smaller than 1.7 Oe at the sample position. Actual data collection started during the following warm up. As shown in Fig. 4(b), upon increasing the temperature, magnetization gradually reduces, and the temperature at which the remnant magnetization vanishes is determined as $T_C$ [8]. All samples are classified into two groups according to the shape of the TRM curves: ferromagnetic-like samples, including Zn-0, Zn-1, and Zn-2, with mean-field-like concave curvature, and non-ferromagnetic samples, including Zn-4 and Zn-8 which show nearly negligible remanent magnetization. Note, that for the ferromagnetic samples, both $T_C$ and TRM decrease upon raising the Zn concentration, and finally vanish for samples Zn-4 and Zn-8. This is not surprising, because such result is expected from the MR data shown in Fig. 3(b): The ferromagnetism is gradually weakened by the doped Zn atoms and finally disappears for the heaviest Zn-doped samples Zn-4 and Zn-8.

To study the evolution of the magnetic properties in all samples, the zero-field-cooling (ZFC) and field-cooling (FC) magnetization measurements under fields of 50 Oe, 100 Oe, 200 Oe, and 500 Oe were carried out, as shown in Fig. 5. As expected, the appearance of the overlap of all zero-field-cooling and field-cooling curves under different fields excludes superparamagnetic or spin-glass behavior, but rather verifies the ferromagnetic feature for the samples Zn-0, Zn-1, as well as Zn-2. Unexpectedly, for sample Zn-4, there is, in addition to the extensively reduced magnetization, an indication of antiferromagnetism, as appearing in Fig. 5(d): all ZFC and FC curves, particularly under 200 Oe, exhibit a maximal value at around 15 K [39]. Interestingly, this is essentially different from (Ga,Mn)As with hole compensation or with much lower Mn concentration, where the reduction of $T_C$ is accompanied with the appearance of a blocking superparamagnetic phase [8, 40, 41]. It is well known, that in (Ga,Mn)As the super-exchange-interaction-induced antiferromagnetic coupling always occurs between the interstitial Mn atom ($Mn_{int}$) and its nearest substitutional Mn atom ($Mn_{sub}$) [1, 16, 42, 43]. The appearance of antiferromagnetism is most probably caused by the increasing fraction of $Mn_{int}$, even comparable to the fraction of $Mn_{sub}$, which is quantitatively analyzed by PIXE and confirmed by first-principles calculations in the later discussion. Such an antiferromagnetic coupling occurs between $Mn_{sub}$ atoms and $Mn_{int}$ occupying the tetrahedral interstitial sites while the exchange between $Mn_{sub}$ atoms and $Mn_{int}$ at the hexagonal position is suppressed [42]. Interestingly, when the Zn implantation fluence increases to $8\times10^{15}$ cm$^{-2}$, the sample Zn-8 exhibits a paramagnetic characteristic, resulting from the continuously increased $Mn_{int}$ fraction in the (Ga,Mn,Zn)As matrix. The appearance of paramagnetism in a so highly *p*-type conductive sample also verifies the theory, that the Mn *d* orbitals of $Mn_{int}$ cannot hybridize with the *p* orbital of the holes in the matrix valence-band and, thus, do not contribute to the hole-mediated ferromagnetism [42, 44]. However the co-doping induced magnetization reduction has been also observed in Be co-doped (Ga,Mn)As prepared by LT-MBE [31, 33].

### D. Substitutional vs. interstitial Mn concentration

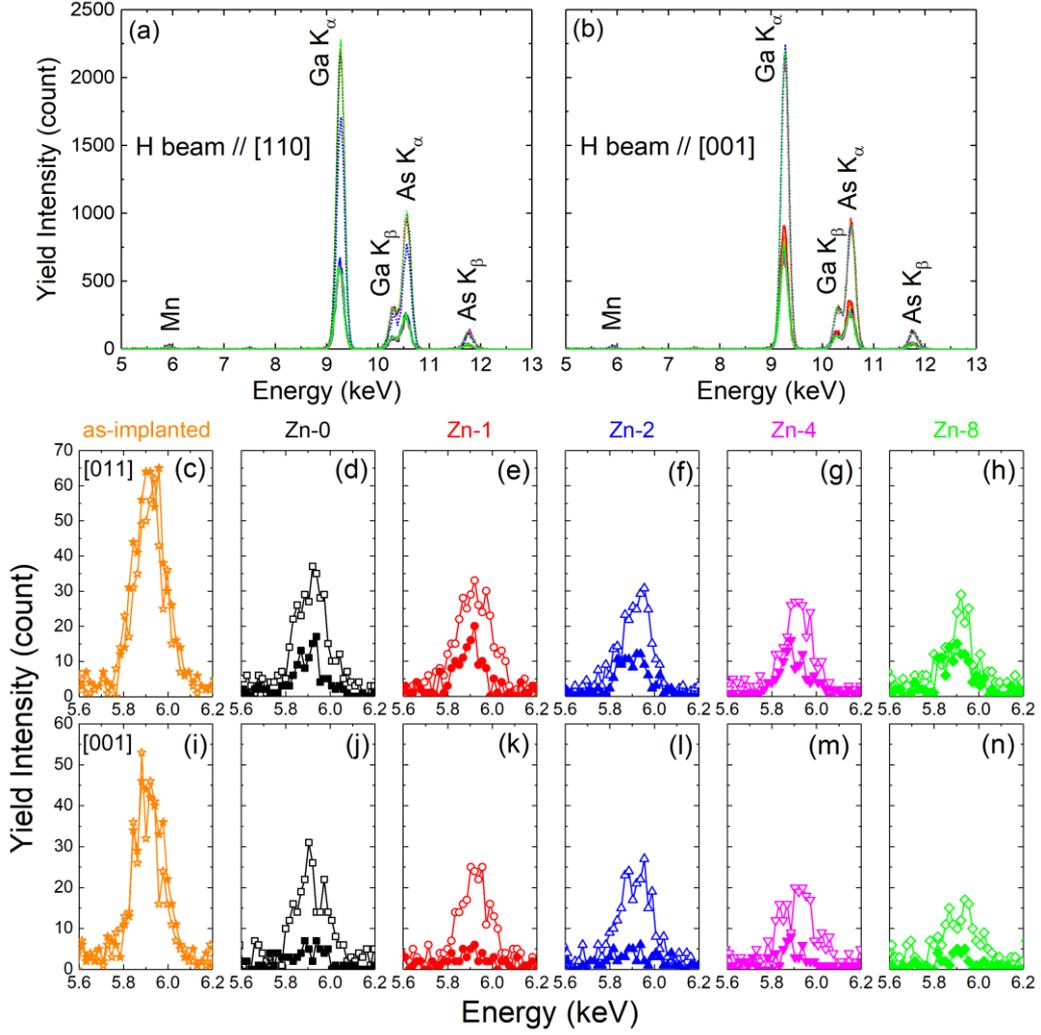

FIG. 6 (Color online) Full particle-induced X-ray emission (PIXE) spectra of all samples if the He beam is aligned along the (a) [011] and (b) [001] directions. Enlarged Mn $K_\alpha$ peak of the PIXE spectra (open scatters: random spectrum and solid scatters: channeling spectrum) for all samples if the He beam is aligned along the (c)(d)(e)(f)(g)(h) [011] and (i)(j)(k)(l)(m)(n) [001] directions. The difference between the yield of these two directions shows that the interstitial Mn can be only observed along the [011] direction.

The quantitative investigation of the Mn lattice localization is carried out by channeling particle-induced X-ray emission (PIXE), due to its high sensitivity to interstitial atoms in the lattice. According to the research by K. M. Yu et al. [31, 33], $Mn_{int}$ at the tetrahedral interstitial sites which contribute to an antiferromagnetic coupling are only detectable, if the He beam is aligned along the [011] direction, however they remain invisible along [001] or [111] directions. Therefore, during our PIXE measurement, all samples were tilted in order to align the He beam parallel to the [011] and [001] directions, respectively, as shown in Fig. 6. Full PIXE spectra at random and channeling conditions are displayed in Figs. 6(a) and (b). The obvious difference between the yield of the random and channeling spectra demonstrates again the epitaxial relation of the recrystallized (Ga,Mn)As:Zn layer on the GaAs substrate, which was already deduced from the TEM and RBS/C results.

We focus now on the Mn $K_\alpha$ peak at an energy of around 5.9 keV, from which the lattice location of Mn can be analyzed in detail. Figs. 6(c) and (i) present the random Mn PIXE signals of the as-implanted sample as open symbols in [011] and [001] directions, respectively. The as-implanted sample is placed as reference to calibrate the Mn concentration, where the Mn concentration is considered to be the same (4.7%), as estimated by SRIM. According to the integral peak intensities of the random spectra, the Mn concentrations in the PLM-treated layers are calculated to be 2.5%, 2.4%, 2.0%, 1.8% and 1.6% for the samples Zn-0, Zn-1, Zn-2, Zn-4, and Zn-8, respectively. The consistence of the relative integral random Mn signal yield along the two directions ([011] and [001]) leads to the conclusion, that upon increasing Zn doping concentration, the Mn atoms gradually escape from the matrix by up to around 36%, and such part of Mn atoms diffuses to the surface and generates an oxide capping layer during the annealing process [45]. For an ideal crystal lattice without any interstitials, the backscattering between the He ions and the impurity atoms occupying the lattice sites is largely suppressed, if the He beam is parallel to a low-index crystal direction, due to the Coulombic potential along the axis, leading to a nearly zero-signal channeling spectrum. However, interstitial atoms intensively increase such backscattering and present a peak in the channeling spectrum, enabling channeling RBS/PIXE to be an element-resolved lattice location technique [33, 34]. As a consequence, the integral yield ratio $\chi_{\min}$ between channeling and random spectra directly responses to the fraction of interstitial impurities, which can be calculated as [32]:

$$f_{\text{int}} = 1 - f_{\text{sub}} = 1 - \frac{1 - \chi_{\min}(\text{Mn})}{1 - \chi_{\min}(\text{Ga})} \quad (1)$$

where $f_{\text{int}}$ and $f_{\text{sub}}$ are the fraction of interstitial and substitutional Mn. Accordingly, the ratio of $\text{Mn}_{\text{int}}/\text{Mn}_{\text{total}}$ gradually increases from 7%, via 15%, 20%, 29% to 35% for the samples from Zn-0 to Zn-8. It is necessary to note, that such a substitutional fraction is calculated from the PIXE spectra if the He beam is aligned along the [011] direction due to the fact that $\text{Mn}_{int}$ is not observable in the channeling spectra obtained along the [001] direction, as shown in Figs. 6 (j) - (n). It is worthy to mention that the secondary absorption and emission were not considered in the PIXE analysis.

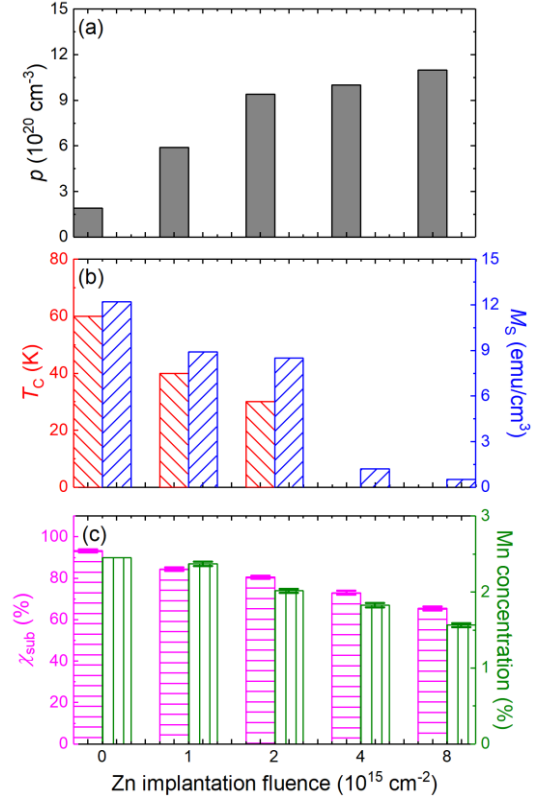

FIG. 7 (Color online) Zn fluence dependence of (a) hole concentration, (b) Curie temperature $T_C$ and saturation magnetization $M_S$ and (c) substitutional fraction $\chi_{sub}$ and the estimated total Mn concentration.

The Zn-fluence-dependent carrier concentration, Curie temperature, saturation magnetization, substitutional fraction, and the Mn concentration are shown in Fig. 7. As presented in Fig. 3, the hole concentration increases till showing saturation characteristics at around $1\times10^{21}$ cm$^{-3}$. As shown in Figs. 6 and 7(c), the raised fraction of Mn$_{int}$ self-consistently explains the non-linearly increased hole concentration: more and more interstitial Mn atoms appear and act as double donors. It is worth noting, that according to the PIXE results, the Zn co-doping does not only result in an increased fraction of Mn$_{int}$ in the layer but also lowers the total Mn content in the layer. Actually, the shift of Mn atoms from substitutional sites to interstitial sites has been theoretically explained [46]. The system energy is increased by the continuous shift of the Fermi level in the valence band, thus will be energetically compensated by the appearance of donors. In our case, the impurity binding energy of Zn in GaAs is ~30 meV [47], which is closer to the valence band maximum (VBM) than Mn (112 meV) [48]. Thus, Mn interstitials appear and compensate holes introduced by the Zn dopant. Our experimental results are supported by first-principles calculations, which will be discussed in Section E.

## E. *Ab initio* calculations

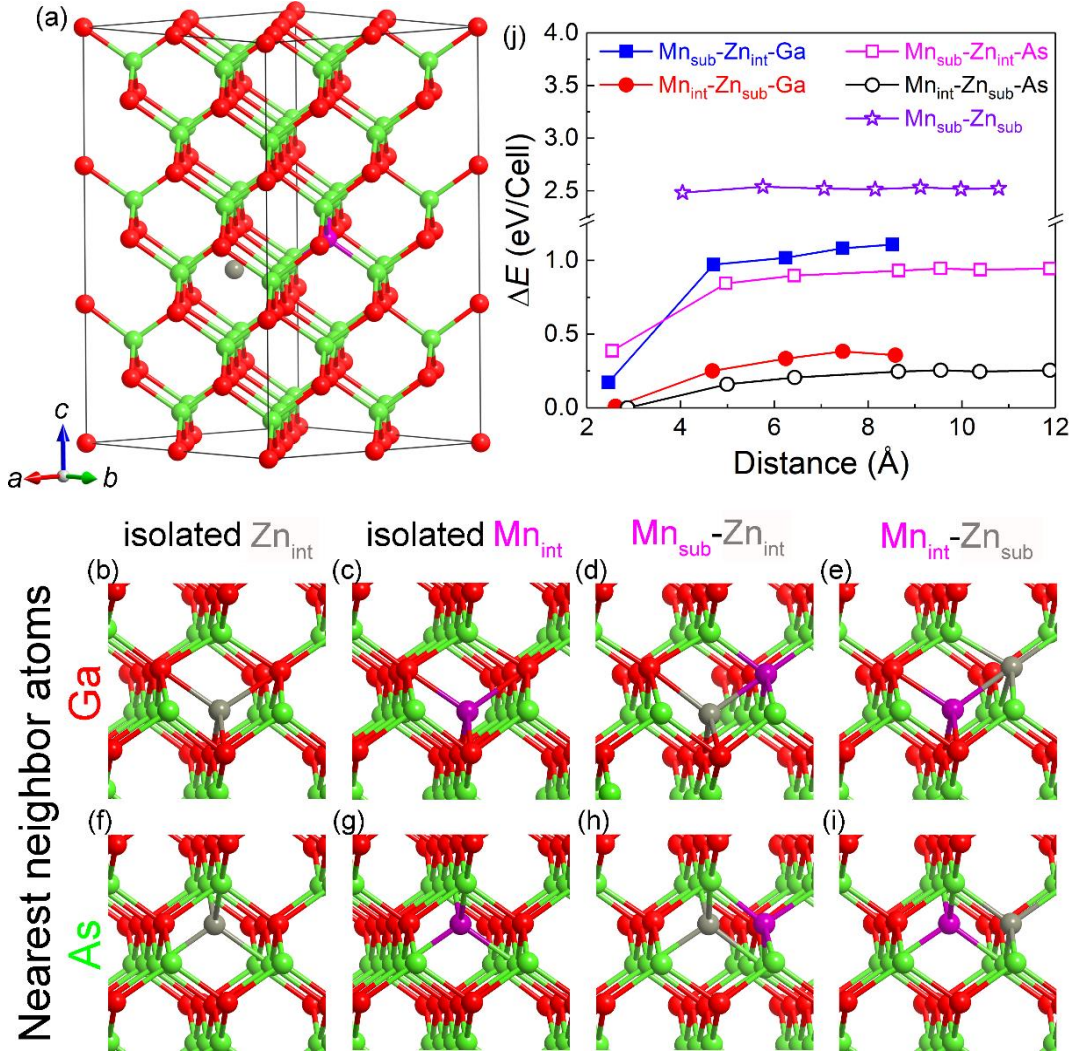

FIG. 8 (Color online) (a) The structure of Zn- and Mn-co-doped GaAs, with (b) (d) (f) (h) substitutional Mn atoms and interstitial Zn atoms and with (c) (e) (g) (i) substitutional Zn atoms and interstitial Mn atoms. The red, violet, green and grey balls refer to Ga, Mn, As and Zn atoms, respectively. The substitutional atoms in (b) (c) (f) (g) are far away from the interstitial ones, so they are not in sight. The interstitial atoms occupy two different sites which are in a tetrahedral position (b to e) with Ga atoms as the nearest neighbors (solid symbols in (j)) or (f to i) with As atoms as nearest neighbors (open symbols in (j)). (j) Upon decreasing the distance between the doped interstitial and substitutional atoms, the distance dependence of cell energy relative to the lowest energy cases, till the generation of dimers between substitutional and interstitial atoms occurs. Case of substitutional Mn and Zn atoms is added for comparison.

Figure 8 presents the *ab initio* computations for the configurations of substitutional Mn atoms together with substitutional Zn atoms, substitutional Mn atoms together with interstitial Zn atoms and substitutional Zn atoms together with interstitial Mn atoms. In the zinc-blende GaAs matrix, there are two different cases for the occupied interstitial atoms where they are in a tetrahedral position with Ga atoms as the nearest neighbors or with As atoms as the nearest neighbors, which are shown in Fig. 8. According to the DFT calculations, it is obvious that the system with interstitial atoms placed closed to As atoms presents lower energy than the one in which interstitial atoms are closer to Ga atoms, for both Mn and Zn. As shown in Fig. 8(j), if the distance between Zn and Mn atoms is larger than 6 Å, the system presents a constant energy for all four situations. However, for both interstitial Mn and Zn atoms, the energy per cell of As-near case is around 0.1 eV lower than the one of Ga-near case. Thus, it is concluded, that the As-near interstitial atoms dominate, which is in excellent agreement with a previous TEM analysis showing that the $Mn_{int}$ preferably occupy As-approach As sites [49, 50]. However, unfortunately, it is impossible to distinguish these two cases by channeling-RBS or PIXE if the He beam is aligned along the [110] direction, because both of them can effectively increase the scattering yield. In good agreement with our RBS/PIXE results, the configurations $Mn_{int}$-$Zn_{sub}$ and $Zn_{int}$-$Mn_{sub}$ show great differences of the cell energy. The former is around 0.7 eV lower than the latter (displayed in Fig. 8j). Together with the results shown in Fig. 6, such a calculation unambiguously confirms the preferential occupation of interstitial sites for Mn atoms but not for Zn atoms. It is worth noting, that the energy sharply decreases if the distance is smaller than 5 angstroms, suggesting the generation of Mn-Zn dimers as shown in Figs. 8(d) (e) (h) (i). Such a phenomenon is reminiscent of Mg and Mn co-doped GaN grown by metal-organic vapor-phase epitaxy (MOVPE) [17]. Noted that the configuration of substitutional Mn atoms together with substitutional Zn atoms presents a much higher energy than the other states. In conclusion, according to the *ab initio* calculation, the substitutional Zn doping prefers to energetically drive the Mn atoms to the interstitial sites, which are close to the As atoms.

## VI. Conclusions

In summary, by ion implantation and subsequent pulsed laser melting, co-doping of Zn is achieved in (Ga,Mn)As, leading to a gradual increase of the hole concentration upon raising the Zn implantation fluence. Thus, ion implantation combined with pulsed laser annealing will be a potential approach to achieve co-doping in DFS materials, while keeping the epitaxial structure. However, unexpectedly, both the Curie temperature and the magnetization exhibit a reduction. Moreover, we observed that the co-doped (Ga,Mn)As to undergo a phase transition from ferromagnetism, via antiferromagnetism, to finally paramagnetism upon increasing co-doping concentration. According to the channeling-RBS/PIXE measurements, we have found that the reduced ferromagnetism and the magnetic phase transition are due to the increased interstitial Mn fraction driven by substitutional Zn doping, which is further confirmed by *ab initio* calculations.

## Acknowledgement

† contribute equally to this work. This research used the resources of Shaheen II at King Abdullah University of Science & Technology (KAUST) in Thuwal, Saudi Arabia. Support by the Ion Beam Center (IBC) at HZDR is gratefully acknowledged. This work is funded by the Helmholtz-Gemeinschaft Deutscher Forschungszentren (HGF-VH-NG-713). The authors Chenhui Zhang, Ye Yuan and Xixiang Zhang thank the support from King Abdullah University of Science and Technology (KAUST), Office of Sponsored Research (OSR), under the award No. OSR-2017-CRG6-3427.The author Chi Xu thanks financial support by Chinese Scholarship Council (File No. 201506680062).

## Reference

[1] T. Dietl, and H. Ohno, Dilute ferromagnetic semiconductors: Physics and spintronic structures, Rev. Mod. Phys. **86**, 187 (2014).
[2] H. Ohno, Making nonmagnetic semiconductors ferromagnetic, Science **281**, 951 (1998).
[3] T. Dietl, H. Ohno, F. Matsukura, J. Cibert, and D. Ferrand, Zener model description of ferromagnetism in zinc-blende magnetic semiconductors, Science **287**, 1019 (2000).
[4] H. Ohno, D. Chiba, F. Matsukura, T. Omiya, E. Abe, T. Dietl, Y. Ohno, and K. Ohtani, Electric-field control of ferromagnetism, Nature **408**, 944 (2000).
[5] D. Chiba, M. Sawicki, Y. Nishitani, Y. Nakatani, F. Matsukura, and H. Ohno, Magnetization vector manipulation by electric fields, Nature **455**, 515 (2008).
[6] D. Chiba, M. Yamanouchi, F. Matsukura, and H. Ohno, Electrical manipulation of magnetization reversal in a ferromagnetic semiconductor, Science **301**, 943 (2003).
[7] T. Dietl, H. Ohno, and F. Matsukura, Hole-mediated ferromagnetism in tetrahedrally coordinated semiconductors, Phys. Rev. B **63**, 195205 (2001).
[8] M. Sawicki, D. Chiba, A. Korbecka, Y. Nishitani, J.A. Majewski, F. Matsukura, T. Dietl, and H. Ohno, Experimental probing of the interplay between ferromagnetism and localization in (Ga, Mn)As, Nat. Phys. **6**, 22 (2009).
[9] T. Dietl, Interplay between Carrier Localization and magnetism in diluted magnetic and ferromagnetic semiconductors, J. Phys. Soc. Jpn. **77**, 031005 (2008).
[10] D. Ferrand, J. Cibert, C. Bourgognon, S. Tatarenko, A. Wasiela, G. Fishman, A. Bonanni, H. Sitter, S. Kolesnik, J. Jaroszyski, A. Barcz, and T. Dietl, Carrier-induced ferromagnetic interactions in *p*-doped $Zn_{1-x}Mn_x$Te epilayers, J. Cryst. Growth **214**, 387 (2000).
[11] D. Ferrand, J. Cibert, A. Wasiela, C. Bourgognon, S. Tatarenko, G. Fishman, T. Andrearczyk, J. Jaroszyński, S. Koleśnik, T. Dietl, B. Barbara, and D. Dufeu, Carrier-induced ferromagnetism in *p*-$Zn_{1-x}Mn_x$Te, Phys. Rev. B **63**, 085201 (2001).

[12] L. Chen, F. Matsukura, and H. Ohno, Electric-field modulation of damping constant in a ferromagnetic semiconductor (Ga,Mn)As, Phys. Rev. Lett. **115**, 057204 (2015).
[13] Y. Ohno, D. Young, B.a. Beschoten, F. Matsukura, H. Ohno, and D. Awschalom, Electrical spin injection in a ferromagnetic semiconductor heterostructure, Nature **402**, 790 (1999).
[14] D. Chiba, F. Matsukura, and H. Ohno, Electric-field control of ferromagnetism in (Ga,Mn)As, Appl. Phys. Lett. **89**, 162505 (2006).
[15] D. Chiba, A. Werpachowska, M. Endo, Y. Nishitani, F. Matsukura, T. Dietl, and H. Ohno, Anomalous Hall effect in field-effect structures of (Ga,Mn)As, Phys. Rev. Lett. **104**, 106601 (2010).
[16] T. Jungwirth, J. Sinova, J. Mašek, J. Kučera, and A.H. MacDonald, Theory of ferromagnetic (III,Mn)V semiconductors, Rev. Mod. Phys. **78**, 809 (2006).
[17] T. Devillers, M. Rovezzi, N.G. Szwacki, S. Dobkowska, W. Stefanowicz, D. Sztenkiel, A. Grois, J. Suffczynski, A. Navarro-Quezada, B. Faina, T. Li, P. Glatzel, F. d'Acapito, R. Jakiela, M. Sawicki, J.A. Majewski, T. Dietl, and A. Bonanni, Manipulating Mn-Mg$_k$ cation complexes to control the charge- and spin-state of Mn in GaN, Sci. Rep. **2**, 722 (2012).
[18] J.F. Ziegler, M.D. Ziegler, and J.P. Biersack, SRIM – The stopping and range of ions in matter (2010), Nucl. Instrum. Meth. B **268**, 1818 (2010).
[19] Y. Yuan, R. Hubner, F. Liu, M. Sawicki, O. Gordan, G. Salvan, D.R. Zahn, D. Banerjee, C. Baehtz, M. Helm, and S. Zhou, Ferromagnetic Mn-implanted GaP: Microstructures vs magnetic properties, ACS Appl. Mater. Interfaces **8**, 3912 (2016).
[20] W. Limmer, A. Koeder, S. Frank, V. Avrutin, W. Schoch, R. Sauer, K. Zuern, J. Eisenmenger, P. Ziemann, and E. Peiner, Effect of annealing on the depth profile of hole concentration in (Ga,Mn)As, Phys. Rev. B **71**, 205213 (2005).
[21] M. Scarpulla, U. Daud, K. Yu, O. Monteiro, Z. Liliental-Weber, D. Zakharov, W. Walukiewicz, and O. Dubon, Diluted magnetic semiconductors formed by ion implantation and pulsed-laser melting, Physica B **340**, 908 (2003).
[22] Y.A. Danilov, H. Boudinov, O. Vikhrova, A. Zdoroveyshchev, A. Kudrin, S. Pavlov, A. Parafin, E. Pitirimova, and R. Yakubov, Formation of the single-phase ferromagnetic semiconductor (Ga,Mn)As by pulsed laser annealing, Phys. Solid Status+ **58**, 2218-2222 (2016).
[23] E. Gan'shina, L. Golik, Z. Kun'kova, G. Zykov, Y.V. Markin, Y.A. Danilov, and B. Zvonkov, Phase Separation in (Ga,Mn)As Layers Obtained by Ion Implantation and Subsequent Laser Annealing, Phys. Solid Status+ **61**, 332 (2019).
[24] O. Dubon, M. Scarpulla, R. Farshchi, and K. Yu, Doping and defect control of ferromagnetic semiconductors formed by ion implantation and pulsed-laser melting, Physica B **376**, 630 (2006).
[25] C. Chen, H. Niu, H. Hsieh, C. Cheng, D. Yan, C. Chi, J. Kai, and S. Wu, Fabrication of ferromagnetic (Ga,Mn)As by ion irradiation, J. Magn. Magn. Mater. **321**, 1130 (2009).
[26] G. Kresse, and J. Furthmuller, Efficient iterative schemes for *ab initio* total-energy calculations using a plane-wave basis set, Phys. Rev. B **54**, 11169 (1996).
[27] P.E. Blochl, Projector augmented-wave method, Phys. Rev. B **50**, 17953 (1994).
[28] J.P. Perdew, K. Burke, and M. Ernzerhof, Generalized gradient approximation made simple, Phys. Rev. Lett. **77**, 3865 (1996).
[29] A. Rushforth, M. Wang, N. Farley, R. Campion, K. Edmonds, C. Staddon, C. Foxon, and B. Gallagher, Molecular beam epitaxy grown (Ga,Mn)(As,P) with perpendicular to plane

magnetic easy axis, J. Appl. Phys. **104**, 073908 (2008).

[30] A.W. Rushforth, N.R.S. Farley, R.P. Campion, K.W. Edmonds, C.R. Staddon, C.T. Foxon, B.L. Gallagher, and K.M. Yu, Compositional dependence of ferromagnetism in (Al,Ga,Mn)As magnetic semiconductors, Phys. Rev. B **78**, 085209 (2008).

[31] K.M. Yu, W. Walukiewicz, T. Wojtowicz, W.L. Lim, X. Liu, U. Bindley, M. Dobrowolska, and J.K. Furdyna, Curie temperature limit in ferromagnetic $Ga_{1-x}Mn_xAs$, Phys. Rev. B **68**, 041308 (2003).

[32] L.C. Feldman, J.W. Mayer, and S.T. Picraux, Materials analysis by ion channeling: submicron crystallography, Academic Press, 2012.

[33] K.M. Yu, W. Walukiewicz, T. Wojtowicz, I. Kuryliszyn, X. Liu, Y. Sasaki, and J.K. Furdyna, Effect of the location of Mn sites in ferromagnetic $Ga_{1-x}Mn_xAs$ on its Curie temperature, Phys. Rev. B **65**, 201303 (2002).

[34] D. Benzeggouta, K. Khazen, I. Vickridge, H. von Bardeleben, L. Chen, X. Yu, and J. Zhao, Quantitative determination of the Mn site distribution in ultrathin $Ga_{0.80}Mn_{0.20}As$ layers with high critical temperatures: A Rutherford backscattering channeling investigation, Phys. Rev. B **89**, 115323 (2014).

[35] J.T. Asubar, S. Sato, Y. Jinbo, and N. Uchitomi, MBE growth and properties of GaMnAs with high level of Zn acceptor incorporation, Phys. Status Solidi A **203**, 2778-2782 (2006).

[36] R. Weiss, and A. Marotta, Spin-dependence of the resistivity of magnetic metals, J. Phys. Chem. Solids **9**, 302 (1959).

[37] H. Nakagawa, J.T. Asubar, Y. Jinbo, and N. Uchitomi, Comparison of annealing effects on Zn-doped GaMnAs and undoped GaMnAs epilayers, Appl. Surf. Sci. **254**, 6648 (2008).

[38] F. Matsukura, H. Ohno, A. Shen, and Y. Sugawara, Transport properties and origin of ferromagnetism in (Ga,Mn)As, Phys. Rev. B **57**, R2037 (1998).

[39] L.M. Schoop, A. Topp, J. Lippmann, F. Orlandi, L. Müchler, M.G. Vergniory, Y. Sun, A.W. Rost, V. Duppel, and M. Krivenkov, Tunable Weyl and Dirac states in the nonsymmorphic compound CeSbTe, Sci. Adv. **4**, eaar2317 (2018).

[40] Y. Yuan, C. Xu, R. Hubner, R. Jakiela, R. Bottger, M. Helm, M. Sawicki, T. Dietl, and S. Zhou, Interplay between localization and magnetism in (Ga,Mn) As and (In,Mn)As, Phys. Rev. Materials **1**, 054401 (2017).

[41] S. Zhou, L. Li, Y. Yuan, A.W. Rushforth, L. Chen, Y. Wang, R. Böttger, R. Heller, J. Zhao, K.W. Edmonds, R.P. Campion, B.L. Gallagher, C. Timm, and M. Helm, Precise tuning of the Curie temperature of (Ga,Mn)As-based magnetic semiconductors by hole compensation: Support for valence-band ferromagnetism, Phys. Rev. B **94**, 075205 (2016).

[42] J. Blinowski, and P. Kacman, Spin interactions of interstitial Mn ions in ferromagnetic GaMnAs, Phys. Rev. B **67**, 121204 (2003).

[43] A.V. Los, A.N. Timoshevskii, V.F. Los, and S.A. Kalkuta, *Ab initio* studies of magnetism in transition-metal-doped silicon carbide, Phys. Rev. B **76**, 165204 (2007).

[44] K. Yu, W. Walukiewicz, T. Wojtowicz, W. Lim, X. Liu, Y. Sasaki, M. Dobrowolska, and J. Furdyna, Determination of free hole concentration in ferromagnetic $Ga_{1-x}Mn_xAs$ using electrochemical capacitance-voltage profiling, Appl. Phys. Lett. **81**, 844 (2002).

[45] M.A. Scarpulla, B.L. Cardozo, R. Farshchi, W.M. Oo, M.D. McCluskey, K.M. Yu, and O.D. Dubon, Ferromagnetism in $Ga_{1-x}Mn_xP$: evidence for inter-Mn exchange mediated by localized holes within a detached impurity band, Phys. Rev. Lett. **95**, 207204 (2005).

[46] J. Mašek, and F. Máca, Self-compensating incorporation of Mn in Ga$_{1-x}$Mn$_x$As, Acta Phys. Pol. A **100**, 315 (2001).

[47] D. Ashen, P. Dean, D. Hurle, J. Mullin, A. White, and P. Greene, The incorporation and characterisation of acceptors in epitaxial GaAs, J. Phys. Chem. Solids **36**, 1041 (1975).

[48] M. Linnarsson, E. Janzén, B. Monemar, M. Kleverman, and A. Thilderkvist, Electronic structure of the GaAs: Mn$_{Ga}$ scenter, Phys. Rev. B **55**, 6938 (1997).

[49] F. Glas, G. Patriarche, L. Largeau, and A. Lemaître, Determination of the local concentrations of Mn interstitials and antisite defects in GaMnAs, Phys. Rev. Lett. **93**, 086107 (2004).

[50] L. Pereira, U. Wahl, S. Decoster, J. Correia, M. da Silva, A. Vantomme, and J. Araújo, Direct identification of interstitial Mn in heavily *p*-type doped GaAs and evidence of its high thermal stability, Appl. Phys. Lett. **98**, 201905 (2011).